\documentclass[aps,prb,twocolumn,superscriptaddress,floatfix]{revtex4-1}
\usepackage[pdftex]{graphicx}
\usepackage{amsmath}
\usepackage{amssymb}
\usepackage{bm}

\begin{document}
\title{Effect of Local Electron-Electron Correlation in Hydrogen-like Impurities in Ge}
\author{H. Sims}
\affiliation{Center for Materials for Information Technology (MINT) and Department of Physics, University of Alabama, Tuscaloosa AL 35487}
\author{E. R. Ylvisaker}
\affiliation{Department of Physics, University of California Davis, Davis CA 95616}
\author{E. \c{S}a\c{s}{\i}o\u{g}lu}
\author{C. Friedrich}
\author{S. Bl\"{u}gel}
\affiliation{Peter Gr\"{u}nberg Institut and Institute for
Advanced Simulation, Forschungszentrum J\"{u}lich and JARA,
52425 J\"{u}lich, Germany}
\author{W. E. Pickett}
\affiliation{Department of Physics, University of California Davis, Davis CA 95616}
\date{\today}
\begin{abstract}
We have studied the electronic and local magnetic structure of the hydrogen interstitial 
impurity at the tetrahedral site in
diamond-structure Ge, using an empirical tight binding + dynamical mean field theory approach
because within the local density approximation (LDA) Ge has no gap. 
We first establish that within LDA the $1s$ spectral density bifurcates
due to entanglement with the four neighboring $sp^3$ antibonding orbitals,
providing an unanticipated richness of behavior in determining under what conditions a local
moment hyperdeep donor or Anderson impurity will result, or on the other hand a gap state
might appear. 
Using a supercell approach, we show that the spectrum, the occupation, and the local moment
of the impurity state displays a strong dependence on the strength 
of the local on-site Coulomb interaction $U$,
the H-Ge hopping amplitude,
the depth of the bare $1s$ energy level $\epsilon_H$, and we address to some extent
the impurity concentration dependence.
In the isolated impurity, strong interaction regime a local moment emerges over
most of the parameter ranges
indicating magnetic activity, and spectral density structure very near (or in) the gap
suggests possible electrical activity in this regime.
\end{abstract}
\maketitle

\section{Introduction}
Due to their importance in electronics technology, isolated defects in 
semiconductors and insulators have a long history.
Low doping levels, arising from isolated shallow defects, provide the
carriers that make semiconductors a dominant technology 
in today's pervasive electronics environment. The primary shallow defects
in the more important semiconductors for most applications (Si, GaAs, Ge)
have been extensively studied, and research turned to
the study of deep levels (states with energies well away from the
edges, deep within the gap). An exploration
by Haldane and
Anderson\cite{haldane} demonstrated, considering intra-atomic repulsion
using the multi-orbital
Anderson impurity in a model semiconductor treated in mean field,
how multiple charge states can arise and be confined within the
semiconducting gap. These charge states will, except accidentally,
be deep levels, and when providing a carrier to the conduction band 
through thermal or electromagnetic excitation, they become deep donor
levels. 

One of the suspected deep donor impurities in semiconductors, and 
seemingly the simplest, is interstitial H in
an elemental semiconductor.  Ge and Si can be prepared ultra-pure, and H possibly
is the most common remaining impurity.  In work that will be 
discussed in more detail later, Pickett, Cohen, and Kittel\cite{PCK} (PCK)
provided evidence that interstitial H produces a {\it hyperdeep donor} level in Ge, with the 
H $1s$ donor state lying not within the gap but perhaps located as deep as 6 eV {\it below}
the gap, near the center of the valence bands.  Their hands-on, self-consistent
mean field treatment in the spirit of correlated band theory (LDA+U) methods leaves
much yet to be decided.

PCK provided a synopsis of the earlier models that had been applied to this
H impurity question. Several H-related defects have been 
observed\cite{Haller1978,complex2,goss,coomer,weber} in
Ge, and most seem to be defect complexes in which H is involved, rather than simply
isolated H impurities. However, local vibrations were observed for isolated H, identified
as (near) bond-centered and in the antibonding or tetrahedral 
sites,\cite{budde,cheney} which is the impurity of interest here. Similar
questions exist for H impurities in the isovalent semiconductors Si\cite{Si_H}
and diamond.

Since that early work, a few model studies have addressed the effects of local interactions
at a single orbital impurity in a semiconducting host.
Yu and Guerrero investigated a one-dimensional Anderson model with an
impurity using the density matrix renormalization group 
approach.\cite{Yu} The strength of the hybridization compared to the
semiconducting gap determined whether the doped-hole density remained
localized near the impurity or instead spread over many sites (25 sites in 
their study). Additional holes were found to be spread throughout the
system, avoiding the impurity region. The H in Ge problem is a physical realization
of the gapped Anderson impurity model (GAIM) studied by Galpin and 
Logan.\cite{Galpin1,Galpin2} They addressed
the GAIM with a self-consistent perturbation theory extended to all
orders, and concluded that for the half-filled case such as we are
in interested in here -- neutral H in undoped Ge -- for any non-zero gap 
the interacting system is
{\it not} perturbatively connected to the non-interacting system.  
This broad claim calls to mind the classic result of Kohn and Majumdar --
separate but related, and with different connotations -- that 
the properties of such a system (in the
non-interacting case) are {\it analytic} in
the strength of a local potential that drives a bound gap state across
the gap edge to become a resonant state in the continuum.\cite{Kohn}

From the earliest electronic structure studies involving H impurities in
Ge, most of the focus has been on defect complexes incorporating H with 
vacancies and other impurities.  Model studies\cite{falicov1,falicov2} gave
way to a number of density functional theory (DFT) based studies; see
Refs. \onlinecite{coomer} and \onlinecite{weber} for representative work. 
DFT studies of isolated H in Ge and other
semiconductors have also been reported,\cite{weber,estreicher,vdWalle,almeida}
giving indications that H provides in Ge a shallow donor or shallow acceptor
depending on its position (see above), or that it is an example of a
negative-U system because of instability of its neutral state. These scenarios,
formulated within a quantum theory of energetics (DFT) but a one-electron
picture of the spectrum,
contrast strikingly with the deep donor possibility posed by PCK. Most of the
existing studies confine their focus to energetics of the H-in-Ge system and on ``energy
level'' positions, without an exposition of the spectral distribution of
the H $1s$ weight.

While the H impurity in an elemental semiconductor is the most primitive realization
of the impurity problem, this type of system has not seen a material-specific treatment of 
the dynamical correlations that will influence its electronic structure and excitation spectrum. 
In this paper we provide results of a dynamical
mean field theory (DMFT) treatment\cite{DMFT1,DMFT2,DMFT3} that sheds light on several of 
the primary issues. 

\section{Method of Calculation}
\subsection{Supercells; host electronic structure}
Interstitial H in intrinsic Ge presents a seemingly simple system: a single
half-filled $1s$ orbital hybridized with a semiconducting bath. 
A neutral H impurity (our interest here) adds one electron that is expected to be
accommodated in an additional ``state'' within the gap or the valence band 
and most likely the latter, since there has been no signature of an 
electrically and magnetically active gap state.   

Anticipating the
disturbance (in density, in screened potential) in an insulator to be localized, we adopt a
supercell representation of the impurity.
We consider a single interstitial hydrogen atom in the tetrahedrally-symmetric antibonding position in
Ge, both in a single (periodic) conventional cubic
diamond-structure supercell (containing 8 Ge atoms, denoted HGe$_8$) and in a $2\times2\times2$ supercell
 of conventional cells (containing 64 Ge atoms and denoted HGe$_{64}$). Choosing two different supercell sizes
allows us to further study the influence of effective H-Ge hybridization beyond simply reducing the hopping
matrix elements and tests to what degree the influence of the H impurity is spatially localized.
There is vibrational evidence\cite{budde,cheney} that
H sits off the tetrahedral site along a $[111]$ direction, thus closer to one of the
four Ge ions than the others, giving it only one Ge nearest neighbor. We do
not treat that possibility here, though the methods we use can be applied to that case. Due to a 
number of uncertainties about materials parameters (and the local density approximation [LDA] 
gap problem due to the small gap of Ge),
we vary the parameters that are not well established, with the goal of 
obtaining a more general picture of the
behavior of a ``H-like'' interstitial in an elemental tetrahedral semiconductor.

One challenge is to deal with the gap underestimation in LDA.  In Ge, the LDA gap is slightly negative,
in contrast to the observed gap of 0.8 eV. Since our objective is an initial
investigation of dynamical correlations at the H site, we adopt the simplest representation of 
the Ge electronic structure. 
Semiconducting Ge will be modeled here using an empirical nearest-neighbor Slater-Koster (S-K) tight-binding
model (eTB) consisting of four Wannier orbitals (one $s$ and three $p$ orbitals) per Ge with parameters obtained
from the work of Newman and Dow.\cite{KNJD} The H-Ge hopping parameters are taken from the
work of Pandey.\cite{pandey}  

\subsection{DMFT parameters}
There are,
inevitably for the current stage of DMFT theory, two parameters that are not known {\it a priori}:
the Coulomb interaction $U$ and the bare on-site $1s$ energy $\epsilon_H$
with respect to the Ge band gap. 
For the single orbital problem there is no Hund's rule $J_H$ interaction
to be concerned about, nor multiplet effects. In fact, for the isolated 
H interstitial the DMFT
result is exact to within numerical uncertainties.  While H-Ge hybridization amplitudes could be
extracted from first-principles DFT calculations, since the gap
problem in LDA leads us to use an eTB model of the Ge electronic structure, we use eTB hopping
amplitudes that were derived in the same spirit.

The hydrogen on-site energy $\epsilon_H$ is varied as part of this
investigation, guided somewhat by the LDA calculations reported in Sec. III.  
Within LDA, where there are no parameters, the $1s$ spectral density for H in the tetrahedral
site unexpectedly bifurcates, so there is no clear point of reference for fixing $\epsilon_H$. 
This splitting is a result of the rather strong hybridization of the $1s$ orbital with the $sp^3$
antibonding orbitals of the four surrounding Ge atoms.  LDA includes, for a localized state
such as a weakly hybridized $1s$ orbital, a spurious self-interaction that raises the LDA site energy
above what is presumed in a LDA+DMFT calculation, providing an extra challenge for determining
$\epsilon_H$.  The H $1s$ orbital likely is not a really strongly
localized state in Ge, but we expect that $\epsilon_H$ = -4 eV should be regarded as upper bound
of the bare $1s$ level. We use the two values -5 eV and -8
eV to span the reasonable range of this parameter.
With regard to the onsite energy, we use the bottom of the gap
as the zero of energy throughout this paper.

The bare ({\it i.e.} unscreened) on-site repulsion $U_0$ for an isolated H $1s$ orbital is 
$\frac{5}{4}$ Ry = 17.01 eV.
This is perhaps surprisingly small for what might seem to be a very small orbital: the $1s$ 
orbital of the smallest atom.  However, it becomes reasonable once it is recognized that the $1s$
radial density $4\pi r^2 \rho(r)$ peaks at 1 $a_0$, whereas the comparable quantity in $3d$
cations peaks at 0.6-0.9 $a_0$ and has $U_0 \approx$ 25-30 eV.  
Screening at a large interstitial site in a small gap insulator is  hard to estimate, with no
comparable values in the literature.  We investigate screened values $U$=7 eV and
$U$=12 eV to span the likely range. 
$U=12$ eV is not much smaller than the unscreened, isolated H value and should allow the
examination of the strong interaction regime.
The choice of 7 eV has specific interest: PCK argued\cite{PCK}
that a lone H $1s$ state would have a bare correlation energy on the order of 1 Ry (our analytic value is
actually 17 eV), and that reduction by screening in an insulator would leave a substantial interaction strength
of 6--7 eV. This amount of reduction, and more, has over the intervening years become commonplace
in understanding the effective values of $U$ in transition metal oxides.

\subsection{Constrained Random-Phase Approximation}
Although we vary both parameters in our impurity Hamiltonian (see Sec.~\ref{subsec:TB}), as well as the H--Ge hybridization,
it is still beneficial to understand the physical value of the interactions in order to both analyze the validity of our range of considered
$U$ and of the predictions of PCK and to motivate and guide future material-specific studies. We do this by employing the
constrained random-phase approximation (cRPA),\cite{cRPA} performed within the full-potential linearized
augmented-plane-wave (FLAPW) method using maximally localized Wannier functions (MLWFs).\cite{cRPA_Sasioglu,Max_Wan}
We use the FLAPW method as implemented in the \texttt{FLEUR} code\cite{fleur} with the PBE exchange-correlation
 potential\cite{PBE} for the ground-state calculations. MLWFs are constructed with the \texttt{Wannier90} code.
\cite{Wannier90,Fleur_Wannier90} The effective Coulomb potential is calculated within the recently developed cRPA method
implemented in the \texttt{SPEX} code \cite{Spex} (for further technical details see Refs.\,\onlinecite{cRPA_Sasioglu},
 \onlinecite{Sasioglu}, and \onlinecite{mpb}). We use a grid of $6\times6\times6$ k-points in our HGe$_8$ cRPA calculations.

The cRPA consists of first writing the polarizability
\begin{widetext}
\begin{align}
P(\mathbf{r},\mathbf{r}^\prime,\omega) = \sum_\sigma\sum_{n}^{\text{occ}} \sum_{m}^{\text{unocc}}
\left[\frac{\psi^{*}_{\sigma{}n}(\mathbf{r})\psi_{\sigma{}m}(\mathbf{r})\psi^{*}_{\sigma{}m}(\mathbf{r}^\prime)
\psi_{\sigma{}n}(\mathbf{r}^\prime)}{\omega-\varepsilon_{\sigma{}m} + \varepsilon_{\sigma{}n} + i\delta} \right.
\left. - \frac{\psi_{\sigma{}n}(\mathbf{r})\psi^{*}_{\sigma{}m}(\mathbf{r})\psi_{\sigma{}m}(\mathbf{r}^\prime)\psi^{*}_{\sigma{}n}
(\mathbf{r}^\prime)}{\omega+\varepsilon_{\sigma{}m} - \varepsilon_{\sigma{}n} - i\delta}\right],
\end{align}
\end{widetext}
where the $\psi_i$ and $\varepsilon_i$ are the DFT wave functions
and  their eigenvalues, and $\sigma$ runs over both spin channels.
If one separates $P$ into $P_l$,
containing the correlated orbitals, and $P_r$, containing the
rest, and if one considers the unscreened Coulomb operator $v$,
one can write\cite{cRPA,cRPA_Sasioglu}
\begin{align}
U =[1-vP_r]^{-1}v
\end{align}
The matrix elements of the effective Coulomb potential $U$ in the
MLWF basis are given by
\begin{align}
U_{\mathbf{R}n_1 n_3;n_4 n_2}(\omega) = \iint
w_{n_1\mathbf{R}}^{*}(\mathbf{r}) w_{n_3\mathbf{R}}
(\mathbf{r}) U(\mathbf{r},\mathbf{r}^{\prime};\omega) \nonumber\\
\times w_{n_4\mathbf{R}}^{*} (\mathbf{r}^{\prime})w_{n_2\mathbf{R}}
(\mathbf{r}^{\prime})\:d^3r\:d^3r^{\prime},
\end{align}
where $w_{n\mathbf{R}}(\mathbf{r})$ is the MLWF at site
$\mathbf{R}$  with orbital index $n$ and $U(\mathbf{r},
\mathbf{r}^{\prime};\omega)$ is calculated within the cRPA.


In our calculations, we choose the Ge $4s4p$ and the H $1s$ orbitals as our
correlated subspace. This is motivated by several considerations. First, we note that,
although only the H $1s$ orbital is treated within DMFT, an interacting picture of the Ge orbitals (taking into account
not only the Hubbard model interactions considered here but other more general and
perhaps non-local terms) is necessary to give the correct gapped band structure.
In our DMFT calculations, this is accomplished through the eTB Hamiltonian, but here it is necessary to
exclude the screening from the Ge $sp$ orbitals in order to get the most accurate assessment of the
screened interactions on the H orbital. The LDA electronic
structure (Fig.~\ref{fig:lda}) in Section~\ref{ssec:lda} makes clear another reason for the necessity of
treating the entire Ge $sp$ + H $s$ subspace within the cRPA: the H 1$s$ state is thoroughly entangled
in the Ge 4$s$ and 4$p$ background. In fact, it appears to be split across two bands, frustrating attempts
to isolate and manipulate it. Naturally, excluding the Ge $sp$ screening increases the resulting value of
the H 1$s$ Hubbard $U$, which can be considered as an upper limit, and so the value of $U$ most
appropriate for our HGe$_8$ DMFT calculations (in which
we only treat the dynamical correlations on the H 1$s$ orbital) is likely smaller than the 11.2 eV we report
in Table~\ref{tab:crpa}. Due to the difference in correlated subspaces considered and the
many varied parameters in the DMFT treatment, it is most appropriate to view this cRPA analysis
as a separate method for understanding the H impurities and as a way to check the
reasonableness of our DMFT approach. We note that the bare (unscreened) value $U_b$ = 16.9 eV is
satisfyingly close to the analytic result for $U_0$ for an isolated H atom (5/4 Ry = 17.01 eV), reflecting the
accuracy of the codes.

\begin{table}[htb]
\begin{ruledtabular}
\begin{tabular}{lcccc}
   Orbital & $U$ & $J$ & $U_b$ & $J_b$  \\ \hline
H  (1\textit{s}) & 11.2&     & 16.9 &     \\
Ge (4\textit{s}) & 7.6 &     & 11.9 &     \\
Ge (4\textit{p}) & 5.7 & 0.3 & 9.2  & 0.4 \\
\end{tabular}
\end{ruledtabular}
\caption[]{cRPA parameters screened Hubbard $U$ and Hund's rule $J$  
($U=F^{0}=\frac{1}{(2l+1)^2}\sum_{m,n}U_{mn;mn}$ and 
 $J=\frac{1}{2l(2l+1)}\sum_{m\neq n}U_{mn;nm}$, 
where $l=0$ and 1 for \textit{s} and \textit{p} orbitals, respectively) 
calculated for H $1s$ and Ge $4s, 4p$ orbitals in HGe$_8$, when all $sp$ transitions
were excluded. When all transitions are excluded we obtain the bare (unscreened) value $U_b$ 
and $J_b$. }\label{tab:crpa}
\end{table}

\begin{table}[htb]
\begin{ruledtabular}
\begin{tabular}{ccccccc}
 & $E_s$ & $E_p$ & $V_{ss\sigma}$ & $V_{sp\sigma}$ & $V_{pp\sigma}$ & $V_{pp\pi}$ \\
\hline
Ge & -5.8 & 1.61 & -1.695 & -2.03 & 2.65 & -0.67 \\
H--Ge & * & --- & -3.30 & 2.16 & --- & --- \\
\end{tabular}
\end{ruledtabular}
\caption[]{Ge S-K empirical tight-binding parameters (in eV) obtained from Newman and Dow,\cite{KNJD} with
H-Ge eTB parameters taken from Pandey.\cite{pandey}  The H $s$ on-site parameter is varied in this 
study, as are the H-Ge hopping parameters; see the text. }\label{tab:SK}
\end{table}

\subsection{Atomic solver}
 We employ a hybridization-expansion continuous-time (CT-HYB)
 quantum Monte Carlo impurity solver,\cite{werner}
 taking advantage of the segment picture\cite{segment} to simplify the computations.
 Our solver is based upon that of the ALPS project;\cite{ALPS_DMFT} we also make use of the ALPS parallel Monte Carlo scheduler.\cite{ALPS_sched}
 Although the CT-HYB solver has many advantages compared to the interaction-expansion method (CT-INT), 
the one-electron self energy calculated from CT-HYB is highly sensitive to Monte Carlo noise. The Dyson equation 
gives the difference between Green's functions obtained from different Monte Carlo simulations,
 preventing the error from canceling. Indeed, the error in the self-energy is proportional to the absolute error
 in the Monte Carlo simulation,\cite{HH} becoming much larger than the actual data even at relatively low frequencies.
 Moreover, one cannot accurately determine other quantities that are sensitive to the 
Green's function and self-energy at all frequencies (such as the occupation of the orbitals).
 Recently, two complementary solutions to this problem have arisen. Boehnke {\it et al.}\cite{LB} showed that,
 by measuring the Green's functions in an orthogonal Legendre basis (limited to a relatively small number of polynomials),
 one can filter out the Monte Carlo noise without losing any accuracy in the computation of 
the Green's functions and self-energies. Hafermann {\it et al.}\cite{HH} derived an 
expression for the self-energy involving a quotient of Green's functions rather than a difference.
 In this formulation, the error in the self-energy is proportional
 to the relative Monte Carlo error, leading to greatly reduced error into high frequencies.
 One can combine these methods for further reduction in the error, and we have implemented both.

\subsection{eTB+DMFT Hamiltonian}\label{subsec:TB}
Our Hamiltonian is that of the Anderson impurity model with a multiorbital ``bath,'' 
which becomes 256 orbitals for our large cell. It can be represented by the following matrix
\begin{equation}
H^\mathbf{k} = \left(\begin{array}{cc}
H_{\text{Ge}}^{\mathbf{k}} & V^{\mathbf{k}} \\
\left(V^{\mathbf{k}}\right)^\dag & H_{\text{imp}}
\end{array}\right)
\end{equation}
where $H_{\text{Ge}}^{\mathbf{k}}$ is the supercell Hamiltonian for Ge 
obtained from the tight-binding model with no additional interaction 
parameters included, $V^{\mathbf{k}}$ is the H-Ge hybridization strength, 
and 
\begin{eqnarray}
H_{\text{imp}} = (\epsilon_{H} - \mu) (\hat n_{1s,\uparrow}+\hat n_{1s,\downarrow})
               + U \hat n_{\uparrow} \hat n_{\downarrow}
\end{eqnarray}
is the hydrogen Hamiltonian. 
There is only a density-density type interaction for a single non-degenerate correlated orbital, 
as required by the segment formulation of the CT-HYB method. 
Real frequency spectra are obtained using the maximum entropy (MaxEnt) method\cite{maxent} as implemented in the ALPS package. 
Static observables such as the average occupation $\langle{}n\rangle{}$, 
double occupation $\langle{}n_\uparrow n_\downarrow\rangle$,
and square of the $z$ component of the spin magnetic moment $\langle{}m_{z}^{2}\rangle$ were measured
during the Monte Carlo simulation ($m_z \equiv n_{\uparrow} - n_{\downarrow})$.

\section{Results and Discussion}
We present here results relating to a H-like impurity in Ge  
as a function of the interaction strength $U$, the magnitude of the H--Ge S-K hopping
amplitude, and $\epsilon_H$, at both a 1:8 and a 1:64 H to Ge ratio. We have considered temperatures ranging 
from $\beta = 5$ eV$^{-1}$ ($T \approx 2300$ K)
to   $\beta = 40$ eV$^{-1}$ ($T \approx 290$ K). One general observation is that the structure of 
the spectra that we obtain does not depend very significantly on temperature, 
so we will neglect temperature
dependence in our discussion. In principle, the existence and character of gap states could show
significant temperature dependence.

\subsection{LDA and $U=0$}\label{ssec:lda}

\begin{figure}[htb]
\includegraphics[]{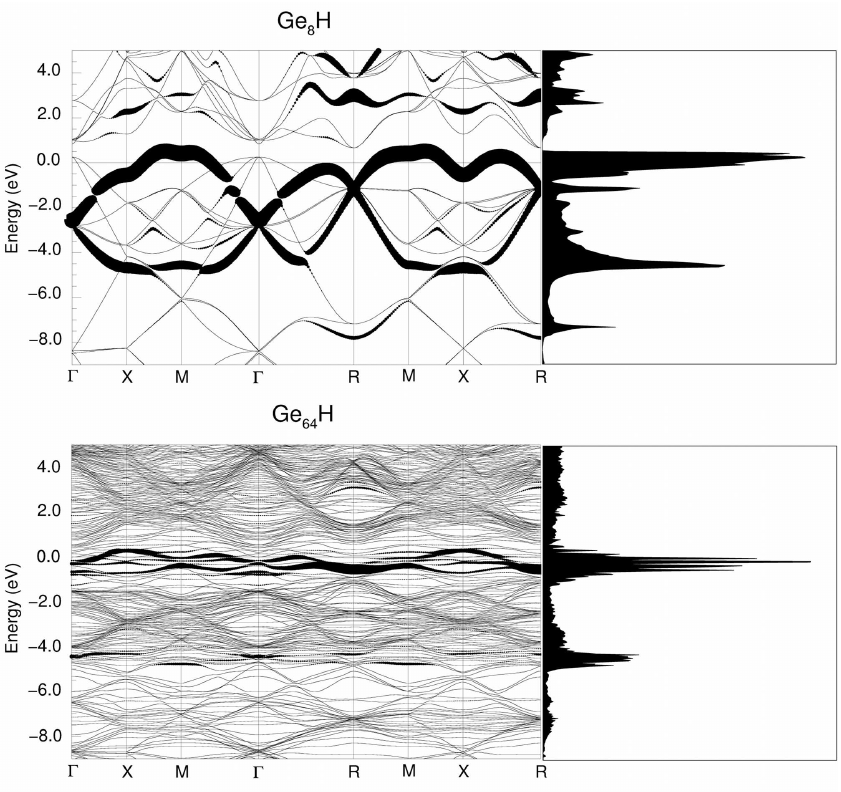}
\caption[]{Left panels: fatbands from LDA emphasizing the H $1s$ character, indicated by the width of the band.
Right panels: H $1s$ projected density of states from the same calculation. For the HGe$_8$ cell
(top panel) seemingly {\it two} $1s$ bands appear, with a total width of nearly 6 eV. The dispersion that is
easy to follow reflects H-H interaction within this small cell.
In the 64 Ge cell (lower panels), the $1s$ spectral density is {\it expelled} 
from the large Ge DOS region (-4 eV to
-1 eV), giving a bifurcation into two peaks
separated by 4-5 eV.}\label{fig:lda}
\end{figure}

As presented in the LDA electronic structure of Fig. \ref{fig:lda} obtained with the  
\texttt{FPLO} code,\cite{fplo} the bands with 
predominantly H $1s$ character can be readily identified. In both the HGe$_8$ and HGe$_{64}$ supercells, 
the H $1s$ local DOS bifurcates into {\it two peaks}, one in and just below the gap and the other around $-5$ eV. 
Whereas these peaks are effectively separate in the larger cell, they form the boundaries of a broad 
but bifurcated $1s$ bandwidth of bandwidth $W\approx$ 5--6 eV. The H spectral density is strongly
expelled by the hybridization $V^{\mathbf{k}}$ from the large Ge DOS region between these peaks, 
leaving no clear way to identify an
on-site H $1s$ energy $\epsilon_H$.

The HGe$_8$ cell result is anomalous in that H appears to introduce {\it two} new bands 
(the system is spin-degenerate)
in the system, whereas there is but a single $1s$ orbital. The substantial dispersion of both bands indicates 
that H interstitials at this concentration are strongly coupled, representing an ordered alloy rather
than an isolated impurity. The band structure is in fact metallic, with the upper $1s$ band partially
filled.  The occurrence of strong $1s$ character in two bands is clarified by the results of the HGe$_{64}$
cell: some of the $1s$ character ($\sim$25\%) of the character lies 4-5 eV below the gap, with the
remained spectral density lying just below the gap and slightly straddling it. The H spectral density 
within LDA is, as noted above, repelled
from the region of large Ge $4p$ DOS, with part going just below and the majority being pushed near the
gap region.  This bifurcation of spectral density may account for the fact that correlated band theory
(the LDA+U method) was unable to produce a single magnetic hyperdeep state\cite{justin} around $\sim$ -5 eV as
would be anticipated from the LDA+U method.
The HGe$_{64}$ results suggest this large cell is effectively in the 
isolated impurity limit.

\begin{figure}[htb]
\includegraphics[]{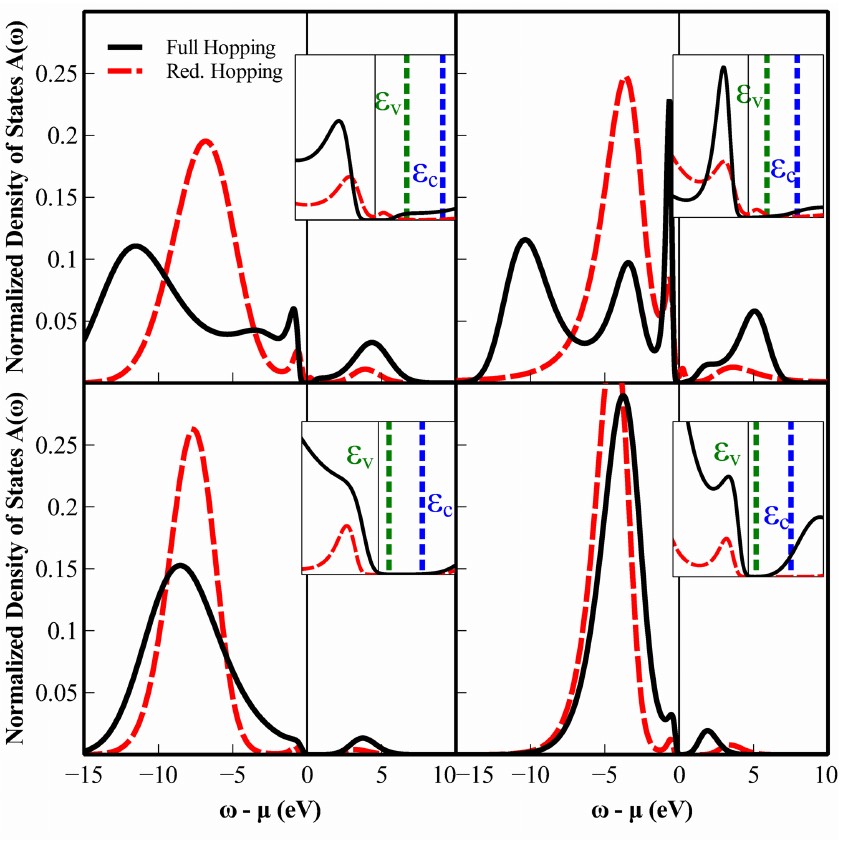}
\caption[]{
The H $1s$ spectral density resulting from eTB+DMFT with $U=0$, {\it i.e.} before 
turning on the interaction. Results are shown for the the two supercells (HGe$_8$, top
panels; HGe$_{64}$, bottom panels), and for 
$\epsilon_H$ = -8 eV (left panels), and -5 eV (right panels).
Reducing the H--Ge hybridization from its full value (full line) leaves a spectrum 
(dashed line) that is dominated by a Gaussian-like
peak centered on $\epsilon_H$. The insets provide an enlargement of the -2 eV to +2 eV
regions, with the Ge band edges indicated by the dotted lines labeled $\epsilon_v$ and $\epsilon_c$.}\label{fig:u0}
\end{figure}

We next survey the non-interacting spectrum (Fig.~\ref{fig:u0}) within our eTB+DMFT picture (using the 
S-K parameters displayed in Table~\ref{tab:SK}). We emphasize that this method is not equivalent to
the LDA results just presented; most notably, in HGe$_{64}$, it contains a gap whereas the LDA bands do not. The gap
in fact is larger than the observed value for Ge, but this allows us to assess more confidently the
tendency toward formation of a gap state in the type of system we are studying: a H-like interstitial
impurity in a Ge-like semiconductor, rather than specifically H in Ge. 
At reduced H--Ge hybridization (dashed red lines in Fig.~\ref{fig:u0}) the spectrum is dominated by a
single Gaussian-like peak centered at
$\epsilon_H$ with only small hybridization effects visible just below and just above the Fermi level, illustrating that the reduced hopping case indeed strongly reduces band structure signatures. 
At full hopping (solid black lines in Fig.~\ref{fig:u0}), the spectrum is substantially spread into a large band with peaks and subpeaks arising
from hybridization between the H $1s$ and Ge $4s$ and $4p$ states, with more of the weight appearing above the gap
and well below $\epsilon_H$. As in the LDA results, the full-hybridization HGe$_8$ calculations show that 
the H $1s$ spectral density is expelled from the region with the largest Ge DOS.
As anticipated, when $\epsilon_H$ is more shallow (-5 eV) the spectral density shows more
weight and structure near the gap. Note that, without magnetic order or strong correlation effects ({\it viz.}
in LDA) the Fermi level must fall within the bands, because our supercells contain an odd number of
electrons which cannot be insulating with spin degeneracy. The insets in Figure~\ref{fig:u0} show the position
of the Ge valence and conduction band edges $\epsilon_v$ and $\epsilon_c$ in the full-hybridization calculations
(we found that the Ge band edges did not depend strongly on the hybridization, so the half-hybridization
values are suppressed for clarity).

The small upward shifts of the Ge bands can be understood by considering the occupancy of the H orbital. Tables~\ref{tab:small_summary} and \ref{tab:large_summary} provide the mean
occupancy $\langle n\rangle$, double 
occupancy $\langle n_{\uparrow}n_{\downarrow}\rangle$ and the mean-square moment $\langle m_z^2\rangle$ 
(which is also the local susceptibility)
for all cases studied. The $1s$ occupation approaches two electrons in the absence
of on-site Coulomb interactions, with the occupation increasing when the impurity level lies deeper and the hybridization
is reduced. When $\langle n\rangle > 1$, all Ge states cannot be occupied as just mentioned above,
so slight hole-doping will occur leading to a weak acceptor picture. Van de Walle and Neugebauer obtained
this type of result\cite{vdWalle} in their LDA studies of isolated H in Ge.
In the smaller cell, the additional electron density is drawn from all Ge atoms. In the large cell, 
charge neutrality is accommodated by relatively small re-organization of electron density on the nearby Ge sites.

\subsection{$U = 7$ eV}

In the absence of guidance from past work, it seemed reasonable to choose interaction strength $U$ values
(reduced from the bare value)  that highlight
likely points of interest in the relationship between impurity energy level, hybridization strength, and on-site interactions.
Choosing $U=7$ eV probes the behavior when $\epsilon_H + U < 0$ as well as when $\epsilon_H + U > 0$, given our
two choices for the $1s$ level. Note that in a mean-field treatment of the interaction, the effective (not bare)
H 1$s$ level would be at $\epsilon_H + U\langle{}n\rangle/2$ (since $\langle{}n_{\uparrow}\rangle$ = $\langle{}n_{\downarrow}\rangle$ = $\langle{}n\rangle/2$).

\begin{figure}[htb]
\includegraphics[]{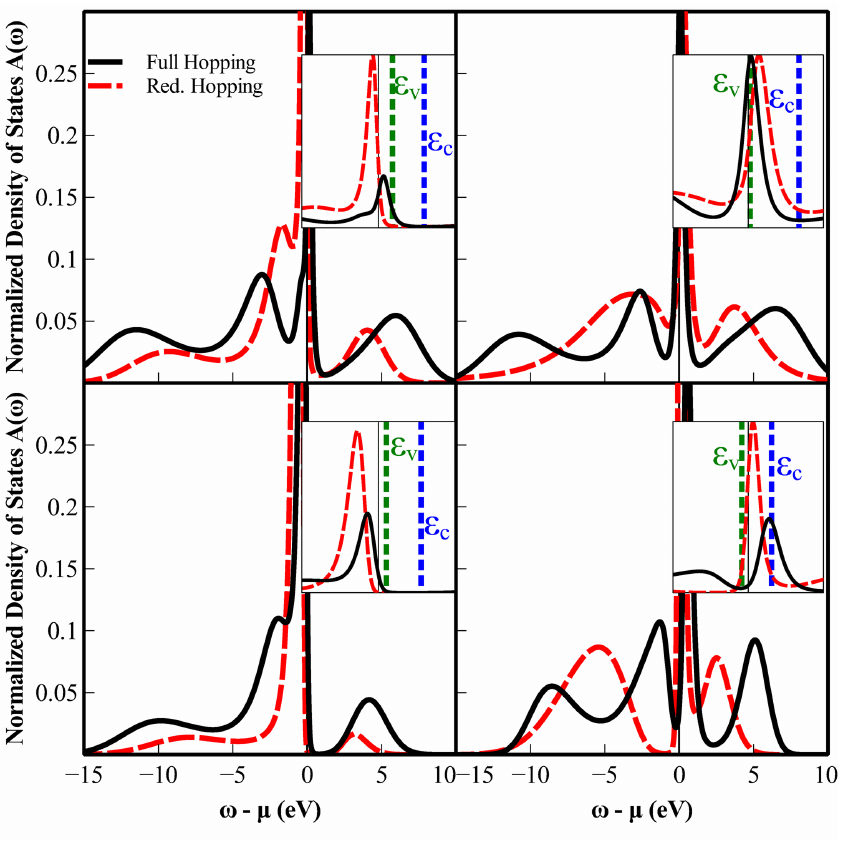}
\caption{The introduction of an on-site Coulomb $U = 7$ eV shifts most of the spectral weight to the vicinity 
of the Ge gap. For $\epsilon_H = -8$ eV (left), two spectral peaks are almost completely occupied, with the 
higher one overlapping a gap state (actual, or incipient) at 0 eV. The peak is  more dramatic near the isolated
impurity limit (lower left) and contains almost all of the $1s$ density. The $\epsilon_H = -5$ eV spectra 
(right) are characterized by a transfer of spectral weight to the Hubbard ``bands'' (now located on either 
side of the gap). (inset) A view of the same MaxEnt spectra between -2 and 2 eV, , with the Ge band edges indicated by the dotted lines labeled $\epsilon_v$ and $\epsilon_c$}\label{fig:u7}
\end{figure}

Figure~\ref{fig:u7} shows the spectral functions at $U=7$ eV, characterized by a large
transfer of the spectral weight from $\epsilon_H$ at U=0 toward the gap region. It is possible that the gap survives
in the $\epsilon_H$ = -8 eV, HGe$_{64}$ spectra, but the imaginary-time and Matsubara Green's functions
(not pictured) suggest that we should believe the small but finite DOS at the Fermi level observed in the MaxEnt data.
This dominant effect of $U$ contrasts
the modest dependence on the H-Ge hopping strength, which is mostly observed in the degree
of splitting in the lower ``Hubbard subbands'' and between the central peak and the upper and lower bands.
This spectrum shift is accompanied by reduced $1s$ occupation as expected 
(see Tables~\ref{tab:small_summary}~and~\ref{tab:large_summary}). At $\epsilon_H$ = -8 eV,
the Ge valence band edge remains above $\mu$ due to the substantial H 1$s$ occupation, with the H gap
state sitting just below $\epsilon_v$ in both supercells. In contrast, the $\epsilon_H$ = -5 eV state
is nearly singly-occupied in both the small and large supercells, which allows $\mu$ to remain at the top of
the Ge valence band. Here, the H gap state sits near one of the band edges, falling squarely in the gap
in HGe$_{64}$.

At full hybridization strength, the H orbital occupation approaches half-filling 
when $\epsilon_H=-5$ eV, but the substantial double occupation
 leaves only a small local moment. For reduced hybridization 
the picture is different. In the $\epsilon_H$=-8 case, a small increase in $1s$ occupation
and a relatively large increase in $\langle{}n_\uparrow{}n_\downarrow\rangle$ yields a decrease
in $\langle{}m_{z}^{2}\rangle$. With a shallower H $1s$ level, however, the orbital remains close to half-filling, and the
$1s$ local moment grows much larger, tending to form a nearly fully-spin-polarized paramagnetic state.
In the HGe$_{64}$ cell, where H-H interaction through the Ge states is negligible, a large local moment appears 
at $\epsilon_H = -5$ at both full and reduced hybridization.

Examining the placement of the gap state, we find that its position appears to be governed by several factors.
It always sits at or near either the top of the Ge valence band or the bottom of the Ge conduction band, depending on which
is closer to the Fermi level. Typically, the gap state sits at the top of the Ge valence band, but in some cases, charge
neutrality dictates that $\mu$ moves into the Ge conduction states. Starting from the Ge band edge, the gap state's
position is further determined by hybridization effects. At full hybridization, the gap state is usually deflected ``upward''
and away from the split lower Hubbard-like sub-bands. However, there is a similar repulsion of spectral weight due to
the upper bands, and so the position is finally determined by the interplay of these factors.

\begin{table*}[htb]
\begin{ruledtabular}
\begin{tabular}{llcccc}
${\bf HGe_8}$ & & \multicolumn{2}{c}{$\bm{\epsilon_H = -5}$ \bfseries{eV}} & 
                 \multicolumn{2}{c}{$\bm{\epsilon_H = -8}$ \bfseries{eV}} \\
 & & Full Hopping & Reduced Hopping & Full Hopping & Reduced Hopping  \\
\cline{3-4} \cline{5-6}
$U=0$ eV & $\langle{}n\rangle$ & 1.66 & 1.89 & 1.79 & 1.94 \\
& $\langle{}n_\uparrow{}n_\downarrow\rangle$ & 0.69 & 0.89 & 0.80 & 0.94 \\
& $\langle{}m_{z}^{2}\rangle$ & 0.28 & 0.11 & 0.19 & 0.06 \\
\hline
$U=7$ eV & $\langle{}n\rangle$ & 1.04 & 1.04 & 1.30 & 1.60\\
& $\langle{}n_\uparrow{}n_\downarrow\rangle$ & 0.21 & 0.12 & 0.38 & 0.60 \\
& $\langle{}m_{z}^{2}\rangle$ & {\bf 0.62} & {\bf 0.80} & {\bf 0.54} & {\bf 0.40} \\
\hline
$U=12$ eV& $\langle{}n\rangle$ & 0.90 & 0.99 & 1.06 & 1.06 \\
& $\langle{}n_\uparrow{}n_\downarrow\rangle$ & 0.11 & 0.04 & 0.18 & 0.08 \\
& $\langle{}m_{z}^{2}\rangle$ & {\bf 0.68} & {\bf 0.91} & {\bf 0.70} & {\bf 0.90} \\
\end{tabular}
\end{ruledtabular}
\caption[]{Local quantities measured during the CT-HYB simulation for the HGe$_8$ supercell. ``Reduced'' 
hopping signifies a tight-binding Hamiltonian in which the H--Ge S-K hopping parameters have been reduced to 
50\% of the value taken from Ref~\onlinecite{pandey}. The double-occupation and local moment show a strong 
dependence on the magnitude of the H--Ge hopping and the position $\epsilon_H$ of the $1s$ level. At half 
hopping, a large local moment arises for $U = 7$ eV in the shallower state, with only a small increase in
$\langle{}m_{z}^{2}\rangle$ as $U$ increases to 12 eV.}\label{tab:small_summary}
\end{table*}

\begin{table*}[htb]
\begin{ruledtabular}
\begin{tabular}{llcccc}
${\bf HGe_{64}}$ & & \multicolumn{2}{c}{$\bm{\epsilon_H = -5}$ \bfseries{eV}} & 
                     \multicolumn{2}{c}{$\bm{\epsilon_H = -8}$ \bfseries{eV}} \\
 & & Full Hopping & Reduced Hopping & Full Hopping & Reduced Hopping  \\
\cline{3-4} \cline{5-6}
$U=0$ eV & $\langle{}n\rangle$ & 1.94 & 1.97 & 1.94 & 1.98 \\
& $\langle{}n_\uparrow{}n_\downarrow\rangle$ & 0.94 & 0.97 & 0.94 & 0.98 \\
& $\langle{}m_{z}^{2}\rangle$ & 0.06 & 0.03 & 0.06 & 0.01 \\
\hline
$U=7$ eV & $\langle{}n\rangle$ & 1.08 & 1.09 & 1.61 & 1.80\\
& $\langle{}n_\uparrow{}n_\downarrow\rangle$ & 0.14 & 0.10 & 0.62 & 0.81 \\
& $\langle{}m_{z}^{2}\rangle$ & {\bfseries 0.80} & {\bfseries 0.89} & {\bfseries 0.37} & {\bfseries 0.18} \\
\hline
$U=12$ eV& $\langle{}n\rangle$ & 0.99 & 1.00 & 1.08 & 1.02 \\
& $\langle{}n_\uparrow{}n_\downarrow\rangle$ & 0.05 & 0.01 & 0.11 & 0.02 \\
& $\langle{}m_{z}^{2}\rangle$ & {\bfseries 0.89} & {\bfseries 0.98} & {\bfseries 0.86} & {\bfseries 0.98} \\
\end{tabular}
\end{ruledtabular}
\caption[]{As in Table~\ref{tab:small_summary}, but for the HGe$_{64}$ supercell. Compared to the smaller cell, 
the local moment state persists even at full H--Ge hybridization and emerges at smaller values of $U$.}
\label{tab:large_summary}
\end{table*}

\subsection{$U = 12$ eV}

Increasing the interaction strength to $U$ = 12 eV, which is near the cRPA value, 
prompts some further spreading of the spectral weight and
additional sharpening (but also shrinking) of the spectral peak in the gap region as the upper Hubbard-like band is pushed well clear.
The five-peak structure that emerges in HGe$_8$ can be understood by the splitting of the upper and lower Hubbard bands due to
H-Ge hybridization (leading to much reduced or absent splitting in the low-hopping case). The $1s$ orbital tends
toward half-filling for all
parameter values reflecting the strong coupling limit. Qualitatively the 1$s$ spectrum in HGe$_8$ is not 
affected greatly by the near-doubling of the
interaction strength. However, in Section~\ref{ssec:lm}, we find that $\langle{}m_{z}^{2}\rangle$ reaches
its maximum value at or near $U$ = 12 eV at all values of $\epsilon_H$ and the H-Ge hybridization
that we consider.

\begin{figure}[htb]
\includegraphics[]{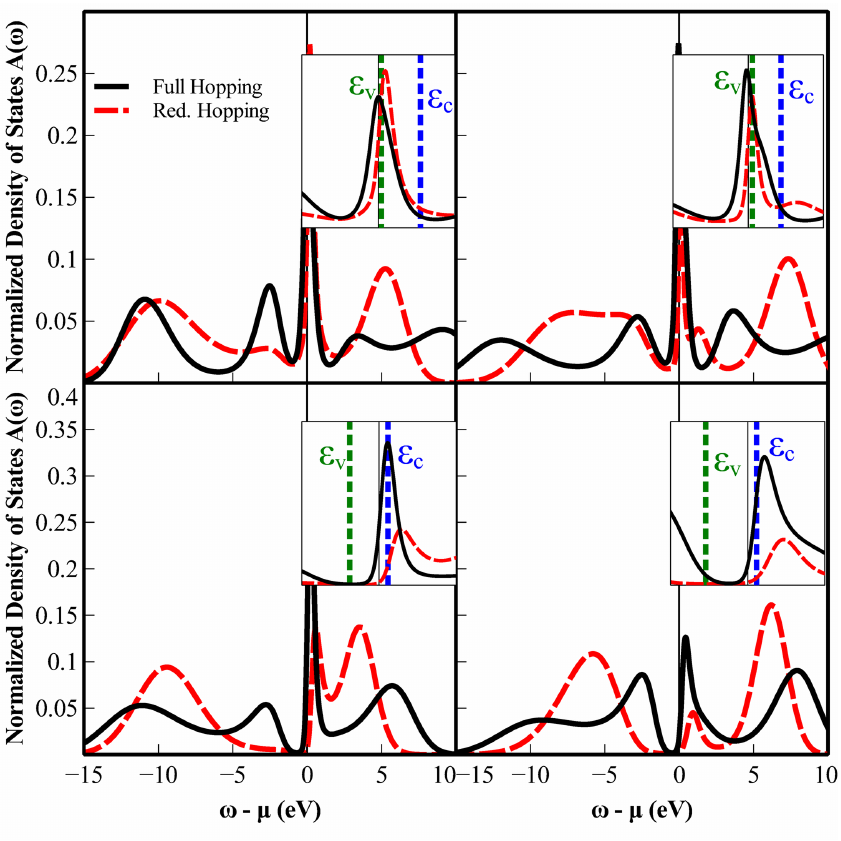}
\caption[]{For strong interactions ($U = 12$ eV), the low-energy quasiparticle-like peak diminishes with respect to the states centered at $\epsilon_H$ and in the conduction band. This effect is most dramatic in the 64-Ge cell, where the spectral weight in the dominant peak at $U=7$ is almost completely transferred to higher energy opening or nearly opening a clear gap. In all cases, the $1s$ orbital approaches half-filling and develops a local moment. (insets) A view of the same MaxEnt spectra between -2 and 2 eV, with the Ge band edges indicated by the dotted lines labeled $\epsilon_v$ and $\epsilon_c$.}\label{fig:u12}
\end{figure}

The H $1s$ spectrum of HGe$_{64}$ begins to differ more strongly from that of the smaller cell when the
interaction becomes strong. A gap that roughly corresponds to the Ge gap but does not necessarily fall
across $\mu$ is restored for both values of the H energy level (Fig.~\ref{fig:u12}).
Further, the sharp peak now just above the gap begins to dissipate, only just surviving in the $\epsilon_H = -8$
eV spectra at full hopping, while nearly disappearing for reduced hopping in the  $\epsilon_H = -5$ spectra.
This behavior suggests that a Mott-Hubbard insulating character of the $1s$ spectrum 
should arise as the impurity limit is approached or as the interaction continues to increase.

\begin{figure}[htb]
\includegraphics[]{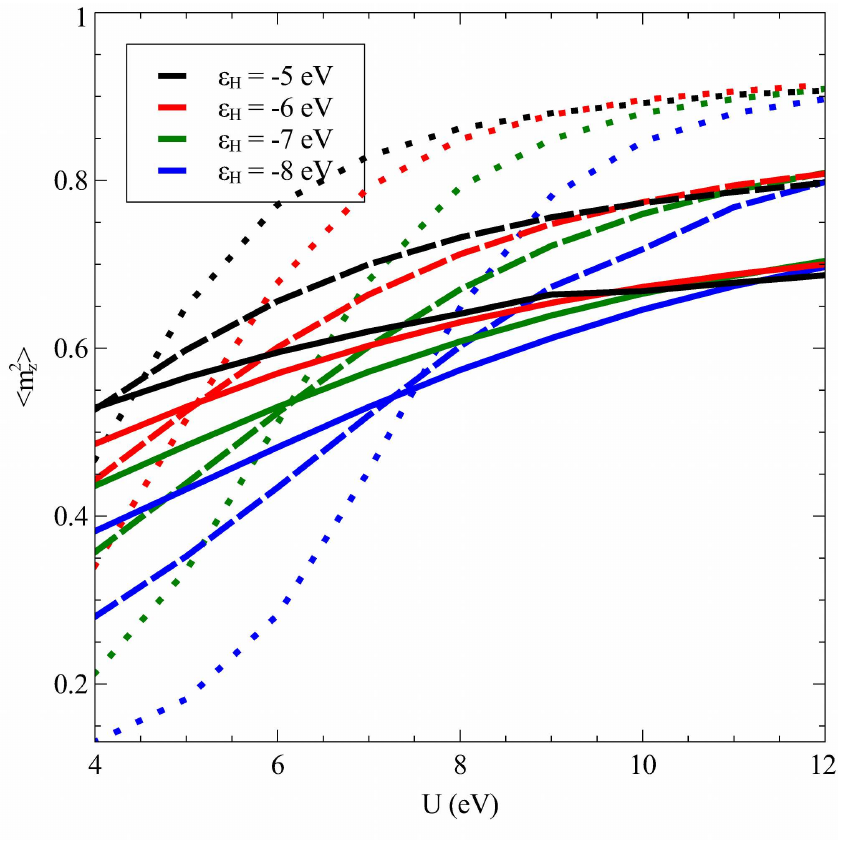}
\caption[]{Monotonic increase of the $1s$ moment on interaction strength $U$, for a range of $1s$ level positions $\epsilon_H$, 
at full H--Ge hopping amplitude (solid), $\frac{3}{4}$ hopping (dashed), and $\frac{1}{2}$ hopping (dotted). 
Reduction in H-Ge hybridization results, as expected, in greater spin-polarization in the 
impurity orbital, but the effect becomes pronounced for reduction below 75\% hopping amplitude 
where the dependence on $\epsilon_H$ also becomes strong. 
}\label{fig:moment}
\end{figure}

\subsection{Transition into the Local Moment State}\label{ssec:lm}

Transitions to a magnetic local moment state $\langle{}m^2_z\rangle \sim$ ${\cal O}$(1) 
occur at different interactions strengths for different values of 
$\epsilon_H$ and H--Ge hopping. To shed further light on this transition, 
we performed a more detailed set of calculations. We varied $U$ between 4 and 12 eV, $\epsilon_H$ between -5 and 
-8 eV, and the H-Ge hopping between 50\% and 100\% of the eTB value, with all calculations performed in the HGe$_8$ 
cell. The results are summarized in Figure~\ref{fig:moment}.

At low $U$, the the local moment at $\frac{1}{2}$ hopping (dotted curve in Figure~\ref{fig:moment})  can be
much smaller than when hybridization is larger. Table~\ref{tab:small_summary} shows that this is due to the
large occupation of the H orbital, particularly when $\epsilon_H$ is below -5 eV. As $U$ increases, the dashed
and dotted curves ($\frac{3}{4}$ and $\frac{1}{2}$ hopping, respectively), show much larger moments, with 
$\langle{}m_{z}^{2}\rangle$ in excess of 0.9 in the H orbital at $\frac{1}{2}$ hopping. This behavior can be understood 
both as a direct result of the hybridization and the increase in interaction strength. The behavior of the curves
at $U = 12$ eV shows that the largest $U$ we considered in the main body of our DMFT study is sufficient
to saturate the local moment: a fortuitous coincidence.
Within each set of curves, 
it can be observed that, until relatively large $U$, the local moment increases with decreasing H level depth. 
Both this trend and its trend toward reversal at very large $U$ arise from the orbital's proximity to half-filling. 
Clearly an orbital at half-filling is able to support a larger moment than one well away from half filling. Below 
$U \approx 8$ eV, only the shallower states are able to push their upper ``Hubbard bands'' above the gap. At large $U$, 
on the other hand, the orbital approaches half-filling regardless of $\epsilon_H$, although for the shallowest $1s$
level, filling falls below unity accompanied by a smaller moment. This tendency to form moments at low effective 
hybridization and large $U$ is consistent with the infinite-$U$, HGe$_{54}$ mean-field treatment of PCK, 
and with lore accumulated in the interim.  In addition, this behavior bears some resemblance to that predicted by 
Li \emph{et al.}\cite{Liarxiv} for impurities on graphene, where there is a vanishing gap and the environment is two-
rather than three-dimensional.

\section{Summary}
We have employed dynamical mean field theory using the CT-HYB solver to study the electronic 
structure of one of the simplest, but still important conceptually and possibly technologically,
impurity systems: an interstitial H-like impurity
in the antibonding interstitial site in a diamond-structure covalent semiconductor representative of
Ge. The non-degenerate $1s$ orbital precludes any necessity of considering Hund's rule magnetic
couping or multiplet effects, so DMFT is the exact solution to the impurity problem in
the limit of large supercell.
Because Ge within LDA has no gap, the non-interacting system 
was represented by an empirical
tight-binding model with parameters based on previous H-Ge studies.  

The H $1s$ spectrum 
shows a rich but systematic behavior as the on-site interaction
parameter $U$, the H-Ge hopping amplitude $V^{\mathbf{k}}$, and the H on-site energy $\epsilon_H$ are varied,
with some parameter ranges possibly giving some insight into Si or diamond hosts as well as Ge. 
The dependence on temperature was studied but found to be minor, and therefore
has not been presented. Our results demonstrate
that electron-electron correlations can play a role in determining the properties of the isolated
H impurity in such a system. If $\epsilon_H + U > 0$, which naively puts the upper ``Hubbard band''
above the gap, the on-site Coulomb repulsion is sufficient to prevent the 
H orbital from acquiring electrons from the surrounding Ge as is the case within LDA. 

The behavior of this system is enriched by the bifurcation, at the LDA level, of the $1s$ spectrum,
with much of the weight in and just below the gap and the rest around 4-5 eV binding energy. Within
the extended tight-binding model we use and with $U$=0, which is not exactly the same as within
LDA but is closely related, the $1s$ occupation is 1.7-1.8 reflecting 
acceptor character (hole-doping) of Ge before
the interaction is turned on.  The distributed spectrum primarily through the valence region may be 
interpreted as the hyper-deep donor character that was envisioned by PCK. 
Increasing $U$ and/or reducing the hybridization moves the $1s$
occupation toward unity -- Anderson impurity character -- with the rate of approach 
affected by the choice of the bare $1s$ site
energy $\epsilon_H$.

To contribute to a more specific study of H in Ge, we have computed the interaction 
parameters $U$ and $J$ in HGe$_8$ for the Ge $sp$ and H $s$ orbitals 
within the constrained RPA formalism. The H 1$s$ Hubbard $U$ is, as expected, much larger than 
those associated with the Ge 4$s$ or 4$p$ states, and is
coincidentally similar to the largest interaction parameter we considered in our DMFT approach. Further studies
are necessary to pin down the behavior of a real H atom in Ge, which will involve
optimization of the energy versus H position together with relaxation of the Ge positions. In
Si, for example, H more commonly assumes a bond-center position although the tetrahedral
interstitial is not far above in energy.\cite{Si_H} 
The methods needed for
those calculations require accurate total energy capability with charge self-consistency,
and must include the weak correlation within the Ge $sp$ bands that gets the band gap
correct as well as the potentially moderately strong correlation associated with the H
impurity.  As such, this H in Ge problem poses a strong challenge for the future. 

\section{Acknowledgments}
H.S. wishes to acknowledge helpful conversations with L. Boehnke and H. Hafermann 
regarding their improved measurement techniques. W.E.P. acknowledges J. C. Smith for discussions of
calculations of H interstitials in (insulating) diamond and xenon. Calculations were performed using computational 
resources from the University of Alabama MINT High Performance Cluster and the NICS Kraken Cray XT5 
under XSEDE project TG-PHY120018. W.E.P. acknowledges support from Department of Energy Stewardship Science
Academic Alliances program under grant DE-FG03-03NA00071 and from the Simons Foundation, and E.R.Y. acknowledges support for 
algorithm development and implementation to the National Science Foundation program Physics at the
Information Frontier through grant PHY-1005503. E.\c{S}, C.F and S.B. acknowledge
the support of DFG through the Research Unit FOR-1346.


\begin{thebibliography}{00}

\bibitem{haldane}F. D. M. Haldane and P. W. Anderson, Phys. Rev. B
  {\bf 13}, 2553 (1976). 
\bibitem{PCK}W. E. Pickett, M. L. Cohen, and C. Kittel, Phys. Rev.  B {\bf 20}, 5050 (1979).

\bibitem{Haller1978} E. E. Haller, Phys. Rev. Lett. {\bf 40}, 584 (1978).

\bibitem{complex2}M. Budde, B. Bech Nielsen, R. Jones, J. Goss, and S. \"Oberg,
  Phys. Rev. B {\bf 54}, 5485 (1996).
\bibitem{goss}J. P. Goss, J. Phys.: Condens. Matter {\bf 15},
  R551 (2003).
\bibitem{coomer}B. J. Coomer, P. Leary, M. Budde, B. Bech Nielsen, R. Jones, S. \"Oberg,
   and P. R. Briddon, Matl. Sci. \& Eng. B {\bf 58}, 36 (1999).
\bibitem{weber}J. Weber, M. Hiller, and E. V. Lavrov, Matl. Sci. Semicond. Processing
   {\bf 9}, 564 (2006).
\bibitem{budde}M. Budde, B. Bech Nielsen, C. Parks Cheney, N. H. Tolk, and L. C. Feldman,
   Phys. Rev. Lett. {\bf 85}, 2965 (2000).
\bibitem{cheney}C. P. Cheney, M. Budde, G. L\"upke, L. C. Feldman, and N. H. Tolk,
   Phys. Rev. B {\bf 65}, 035214 (2002).
\bibitem{Si_H} E. A. Davis, J. Non-Crystall. Solids {\bf 198-200}, 1 (1996).

\bibitem{Yu}C. C. Yu and M. Guerrero, Phys. Rev. B {\bf 54},
  8556 (1996).
\bibitem{Galpin1}M. R. Galpin and D. E. Logan, Phys. Rev. B
  {\bf 77}, 195108 (2008).
\bibitem{Galpin2}M. R. Galpin and D. E. Logan, Eur. Phys. J. B {\bf 62}, 129 (2008).
\bibitem{Kohn}W. Kohn and C. Majumdar, Phys. Rev. {\bf 138}, A1617 (1965).

\bibitem{falicov1}J. Oliva and L. M. Falicov, Phys. Rev. B {\bf 28}, 7366 (1983).
\bibitem{falicov2}J. Oliva, Phys. Rev. B {\bf 29}, 6846 (1984).

\bibitem{estreicher}S. K. Estreicher and Dj. M. Maric, Phys. Rev. Lett. {\bf 70},
  3963 (1993).
\bibitem{vdWalle}C. G. van de Walle and J. Neugebauer, Nature (London) {\bf 423},
   626 (2003).
\bibitem{almeida}L. M. Almeida, J. Coutinho, V. J. B. Torres, R. Jones, and
   P. R. Briddon, Matl. Sci. Semicond. Processing {\bf 9}, 503 (2006).


\bibitem{DMFT1} E. M\"uller-Hartmann, Z. Phys. B {\bf 74}, 507 (1989).
\bibitem{DMFT2} W. Metzner and D. Vollhardt, Phys. Rev. Lett. {\bf 62}, 324 (1989).
\bibitem{DMFT3} A. Georges and G. Kotliar, Phys. Rev. B {\bf 45}, 6479 (1992).

\bibitem{KNJD} K. E. Newman and J. D. Dow, Phys. Rev. B {\bf 30}, 1929 (1984).

\bibitem{pandey} K. C. Pandey, Phys. Rev. B {\bf 14}, 1557 (1976).

\bibitem{cRPA}
F. Aryasetiawan, M. Imada, A. Georges, G. Kotliar, S. Biermann, and A.
I. Lichtenstein, Phys. Rev. B \textbf{70}, 195104 (2004); F.
Aryasetiawan, K. Karlsson, O. Jepsen, and U. Sch\"{o}nberger,
Phys. Rev. B \textbf{74}, 125106 (2006); T. Miyake, F.
Aryasetiawan, and M. Imada Phys. Rev. B \textbf{80}, 155134
(2009); B-C. Shih, Y. Zhang, W. Zhang, and P. Zhang, Phys. Rev. B
\textbf{85}, 045132 (2012); E. \c{S}a\c{s}{\i}o\u{g}lu, C. Friedrich, 
and S. Bl\"{u}gel, Phys. Rev. Lett. \textbf{109}, 146401 (2012); H. Sims, 
W. H. Butler, M. Richter, K. Koepernik, E. \c{S}a\c{s}{\i}o\u{g}lu, C. Friedrich, 
and S. Bl\"{u}gel, Phys. Rev. B \textbf{86}, 174422 (2012).



\bibitem{cRPA_Sasioglu}
E. \c{S}a\c{s}{\i}o\u{g}lu, C. Friedrich, and S. Bl\"{u}gel, Phys.
Rev. B \textbf{83}, 121101(R) (2011).

\bibitem{Max_Wan}
N. Marzari and D. Vanderbilt, Phys. Rev. B \textbf{56}, 12847
(1997).

\bibitem{fleur}
http://www.flapw.de

\bibitem{PBE}
J. P. Perdew, K. Burke, and M. Ernzerhof, Phys. Rev. Lett. {\bf
77}, 3865 (1996).

\bibitem{Fleur_Wannier90}
F. Freimuth, Y. Mokrousov, D. Wortmann, S. Heinze, and S.
Bl\"{u}gel, Phys. Rev. B \textbf{78}, 035120 (2008).


\bibitem{Wannier90}
A. A. Mostofi, J. R. Yates, Y.-S. Lee, I. Souza, D. Vanderbilt,
and N. Marzari, Comput. Phys. Commun. \textbf{178}, 685 (2008).

\bibitem{Spex}
C. Friedrich, S. Bl\"{u}gel and A. Schindlmayr, Phys. Rev. B.
\textbf{81}, 125102 (2010).

\bibitem{Sasioglu}
E. \c{S}a\c{s}{\i}o\u{g}lu, A. Schindlmayr, C. Friedrich, F.
Freimuth and S. Bl\"{u}gel, Phys. Rev. B. \textbf{81}, 054434
(2010).

\bibitem{mpb}
C. Friedrich, A. Schindlmayr, and S. Bl\"{u}gel, Comp. Phys. Comm.
{\bf 180}, 347 (2009).

\bibitem{werner} P. Werner and A. J. Millis, Phys. Rev. B {\bf 74}, 155107 (2006).

\bibitem{segment} P. Werner, A. Comanac, L. de'Medici, M. Troyer, and A. J. Millis, Phys. Rev. Lett. {\bf 97}, 076405 (2006).

\bibitem{ALPS_DMFT} E. Gull, P. Werner, S. Fuchs, B. Surer, T. Pruschke, and M. Troyer, Computer Physics Communications {\bf 182}, 1078 (2011).

\bibitem{ALPS_sched} M. Troyer, B. Ammon, and E. Heeb, Lect. Notes Comput. Sci. {\bf 1505}, 191 (1998).

\bibitem{HH} H. Hafermann, K. R. Patton, and P. Werner, Phys. Rev. B {\bf 85}, 205106 (2012).

\bibitem{LB} L. Boehnke, H. Hafermann, M. Ferrero, F. Lechermann, and O. Parcollet, Phys. Rev. B {\bf 84}, 075145 (2011).

\bibitem{maxent}  R. N. Silver, D. S. Sivia, and J. E. Gubernatis, Phys. Rev. B {\bf 41}, 2380 (1990);
M. Jarrell and O. Biham, Phys. Rev. Lett. {\bf 63}, 2504 (1989).

\bibitem{fplo} K. Koepernik and H. Eschrig, Phys. Rev. B {\bf 59}, 1743 (1999); http://www.fplo.de

\bibitem{justin}J. C. Smith and W. E. Pickett, unpublished.

\bibitem{Liarxiv} C. Li, J.-X. Zhu, and C.S. Ting, arXiv:1106.5827v1.


\end{thebibliography}
\end{document}